\documentclass{article}
\usepackage{spconf,amsmath,graphicx}
\usepackage{amsfonts}
\usepackage{enumitem}
\setlist{nosep, leftmargin=14pt}

\usepackage{mwe} 

\title{Enhancing non-mass Breast Ultrasound Cancer Classification with Knowledge Transfer}
\name{Yangrun Hu$^{1,3}$\sthanks{These authors contribute equally to this work.}, Yuanfan Guo$^{1,3*}$, Fan Zhang$^{2}$, Mingda Wang$^{1,3}$, Tiancheng Lin$^{1,3}$, Rong Wu$^{2}$, Yi Xu$^{1,3}$\sthanks{Corresponding Author}}
\address{$^{1}$Shanghai Key Lab of Digital Media Processing and Transmission, Shanghai Jiao Tong University\\  $^{2}$Department of Ultrasound, Shanghai General Hospital, Shanghai Jiao Tong University School of Medicine \\$^{3}$MoE Key Lab of Artificial Intelligence, AI Institute, Shanghai Jiao Tong University } 
\begin{document}
%
\maketitle
\begin{abstract}

Much progress has been made in the deep neural network (DNN) based diagnosis of mass lesions breast ultrasound (BUS) images. However, the non-mass lesion is less investigated because of the limited data. Based on the insight that mass data is sufficient and shares the same knowledge structure with non-mass data of identifying the malignancy of a lesion based on the ultrasound image, we propose a novel transfer learning framework to enhance the generalizability of the DNN model for non-mass BUS with the help of mass BUS. Specifically, we train a shared DNN with combined non-mass and mass data. With the prior of different marginal distributions in input and output space, we employ two domain alignment strategies in the proposed transfer learning framework with the insight of capturing domain-specific distribution to address the issue of domain shift. Moreover, we propose a cross-domain semantic-preserve data generation module called CrossMix to recover the missing distribution between non-mass and mass data that is not presented in training data. Experimental results on an in-house dataset demonstrate that the DNN model trained with combined data by our framework achieves a 10\% improvement in AUC on the malignancy prediction task of non-mass BUS compared to training directly on non-mass data.

\end{abstract}

\section{Introduction}
\label{intro}

\begin{figure}[tb]
\centering
\setlength{\abovecaptionskip}{-10pt}
\setlength{\belowcaptionskip}{-50pt}
\includegraphics[width=0.35\textwidth]{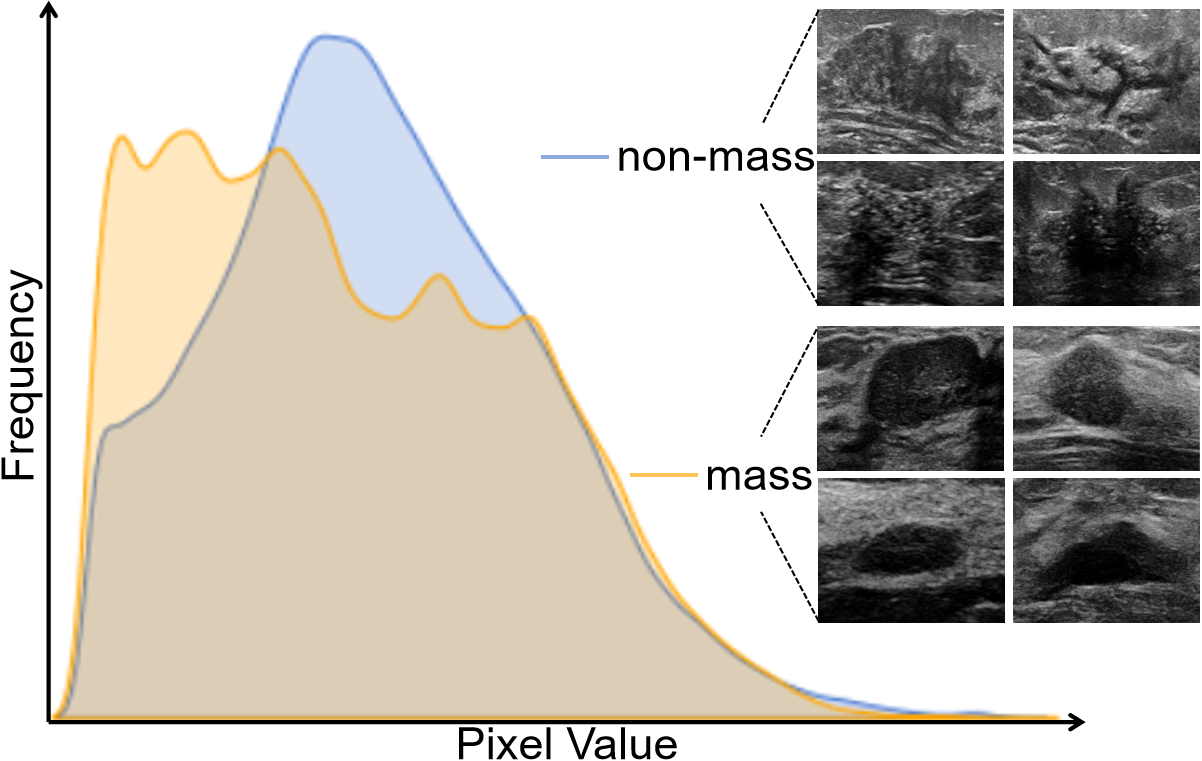}
\caption{The example of input domain shift with the average pixel distribution and the sample images of mass and non-mass breast lesion ultrasound images.}
\label{PD}
\end{figure}

\begin{figure*}[tb]
\setlength{\belowcaptionskip}{-5pt}
\setlength{\abovecaptionskip}{-15pt}
\centering
\includegraphics[width=0.9\textwidth]{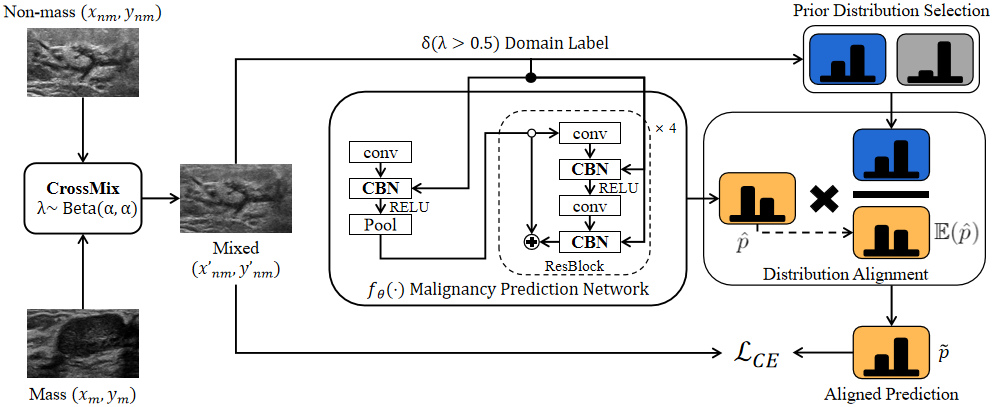}
\caption{The proposed transfer learning framework with CBN, DA and CrossMix.}
\label{framework}
\end{figure*}
Breast cancer is one of the most common cancers and the leading cause of death for women worldwide. The early and accurate diagnosis of breast cancer is an essential task for imagining examination. Breast Ultrasound (US) is a widely adopted imaging modality for early breast cancer diagnosis with the advantages of being non-invasive, safe, and relatively inexpensive~\cite{ultrasound_advantage}. 
To reduce the workload of radiologists and improve diagnostic accuracy, the deep-learning-based computer-aided diagnosis (CAD) system has been developed to help radiologists in breast cancer Benign/Malignant classification~\cite{ultrasound_BM1,ultrasound_BM2}. These works mainly focus on mass breast ultrasound images. However, besides breast masses, we often encounter some non-mass breast lesions (NMLs) in clinical work which demonstrate the space-occupied effect in US. Compared to mass lesions, the morphological feature between benign and malignant NMLs in US are more overlapped, making it difficult for radiologists to make a correct judgment. Therefore, it is very important to develop the CAD system for non-mass breast lesions.

However, because of the limited data~\cite{choi2016additional,zhang2018non}, it is hard for a deep-learning-based model to draw the complete data distribution and it is challenging to train a robust model that could generalize to the real-world application. Nevertheless, mass lesion data is often sufficient to recover the true distribution. Therefore, with the same knowledge structure of learning to capture the relationship between ultrasound image representation of a tumor and its malignancy, we are interested in recovering data distribution of non-mass data leveraging knowledge transferred from mass data. A naive approach is to train a malignancy prediction model with combined non-mass and mass data. However, as illustrated in Fig~\ref{PD}, there is a domain shift in input space, represented as a shift on pixel value distribution. In addition, as indicated in previous research~\cite{xiang20213}, the malignancy rate of the non-mass lesion is different from the mass one, which could be summarized as domain shift in output space. Conventional machine learning algorithms and modern neural networks often adapt poorly to domain shift~\cite{sun2016return}. Therefore, to address the issue of domain shift in input space, we propose to employ conditional batch normalization (CBN)~\cite{CBN} with the insight of modulating statistical features for each domain specifically and thus rescuing the model from adapting the input statistics. As for output domain shift, inspired by recent works in domain adaptation~\cite{AdaMatch}, we propose a clinical prior guided distribution alignment (DA) module to align the malignancy prediction distribution with the ground truth distribution of each domain specifically.
Finally, based on the underlying assumptions that the mixed non-mass and mass images with the same semantic label can help to recover the joint distribution of the non-mass image and its corresponding semantic class (i.e. benign/malignant), we propose CrossMix, a cross-domain data generation module, to generate linearly combined non-mass and mass images that could significantly enhance the diversity of training data. Overall, our contributions are summarized as:
\begin{itemize}
    \item We propose a novel transfer learning framework consisting of domain alignment and a cross-domain data generation module to enhance the learning of non-mass breast lesion ultrasound images with limited training data size with the knowledge transferred from mass lesion.
    \item The experimental results show that with the proposed transfer learning framework, we could achieve a significant improvement on the malignancy prediction performance (10\% on AUC) for non-mass breast lesion data.
\end{itemize}

\section{Problem Definition and Preliminaries}
\label{preliminary}

Some notations and preliminaries in this research need to be clearly defined before introducing the proposed transfer learning framework. Generally, given a source domain $\mathcal{D}_t=\{\mathcal{X}_s, P(X_s)\}$ and its corresponding task $\mathcal{T}_s = \{ \mathcal{Y}_s, P(Y_s|X_s)\}$, transfer learning is the process of improving the predictive function $f_\theta(\cdot): \mathcal{X}_t \to \mathcal{Y}_t$ on target domain $\mathcal{D}_t = \{ \mathcal{X}_t, P(X_t)\}$ and its corresponding task $\mathcal{T}_t = \{ \mathcal{Y}_t, P(Y_t|X_t)\}$ by using related information from $\mathcal{D}_s$ and $\mathcal{T}_s$~\cite{pan2009survey}. Specifically in this research, the source domain is mass data and the target domain is non-mass data. It is clear that $\mathcal{X}=\mathcal{X}_s = \mathcal{X}_t$ (both are ultrasound images) and $\mathcal{Y}=\mathcal{Y}_t = \mathcal{Y}_s=\{0,1\}$ (0 is benign and 1 is malignant). Therefore, a basic transfer learning framework could naively be represented as learning a shared predictive function $f_\theta (\cdot)$ combining two domains. However, the clinical prior indicates the existence of domain shift where $P(X_s) \neq P(X_t)$ since the pixel distribution is different among non-mass an mass images and $P(Y_s) \neq P(Y_t)$ since the malignancy rate is different between non-mass and mass data. Domain shift has been proved empirically by previous researches~\cite{CBN,AdaMatch} to bring non-negligible negative effects on the performance of a learner on the target domain. 

\section{Method}

With the challenges and assumptions introduced above, an overview of our proposed transfer learning framework could be seen in Fig~\ref{framework}. The framework mainly consists of: (1) a shared malignancy prediction function $f_\theta (\cdot): \mathcal{X} \to \mathcal{Y}$ instantiated as a neural network parameterized by $\theta$; (2) CBN and DA modules for domain alignment in input and output space; (3) a cross-domain data generation module CrossMix to recover the missing distribution in non-mass data with the help of mass data.

\subsection{Domain Alignment}
\label{alignment}
It is common for modern neural networks to employ batch normalization to capture the statistics feature of input data. However, when $P(X_s) \neq P(X_t)$, BN has been proved significantly affected by domain shift~\cite{AdaBN}. 
To address this issue, a straightforward solution is to capture domain-specific input statistics such that BN learns statistics based on data from a similar distribution. To this end, we are interested in conditional batch normalization (CBN) proposed by Vries et.al~\cite{CBN}. The key insight of CBN is to modulate the statistics in the BN layer by predicting the scaling factor $\gamma$ and bias factor $\beta$ for each domain with a one-hidden-layer MLP in a residual manner. Specifically, denoting domain label (mass/non-mass) as $e$, then an MLP is employed before BN to predict residuals of $\gamma$ and $\beta$ as:
\begin{equation}
    \Delta \beta, \Delta \gamma = MLP(e).
\end{equation}
Then, these residual terms are added to the original factors as:
\begin{equation}
    \hat{\beta} = \beta + \Delta \beta \quad \hat{\gamma} = \gamma + \Delta \gamma,
\end{equation}
where $\hat{\beta}$ and $\hat{\gamma}$ are used as bias and scaling factors for the following BN. 

Apart from the input shift, the output domain shift would also bring negative effects on the learning of the classification head of the neural network. With the similar insight of alignment through modulating domain-specific information, DA aims to align the prediction distribution to the prior distribution of non-mass and mass data, respectively. Formally, given the model prediction of the malignancy score $\hat{p}_m = f_\theta(x_m)$ of a mass lesion image $x_m$, the aligned malignancy score $\tilde{p}_m$ is:
 \begin{equation}
     \tilde{p}_m = \text{normalize}( \hat{p}_m \times  \frac{q_m}{\mathbb{E}({\hat{p}_m})}),
 \end{equation}
 where $q_m$ is a prior malignancy rate implemented as the percentage of malignant mass data in the training data, and in practice $\mathbb{E}({\hat{p}_m})$ is implemented by the moving average of the model's prediction on training data. $\text{normalize}(\cdot)$ is used to scale the aligned prediction to in range $[0, 1]$. Similarly, for non-mass data we have:

\begin{equation}
\tilde{p}_{nm} = \text{normalize}( \hat{p}_{nm} \times  \frac{q_{nm}}{\mathbb{E}({\hat{p}_{nm}})}).
\end{equation}

 

\subsection{CrossMix}
\label{crossmix}
With domain alignment, knowledge transfer is feasible by training a shared neural network with mass and non-mass data. However, the limited non-mass training data makes it difficult for the network to recover the true data distribution. Typical data augmentation via geometric transformation could partially fix this problem. However, with the assumption of $\mathcal{X}_s=\mathcal{X}_t$ and $P(X_s) \neq P(X_t)$, it is obvious that the training data is not able to cover samples that are between the distribution of non-mass and mass data. With the principle of Vicinal Risk Minimization (VRM)~\cite{VRM}, it is clear that given a mixed image of malignant non-mass and mass image, the radiologist would still recognize it as malignant. Following VRM, we could generate cross-domain images that share the same semantic labels to further recover the missing part of the ground truth joint distribution. Inspired by an easy yet efficient implementation of such a principle in the single domain scene called MixUp~\cite{MixUp}, we propose CrossMix to generate synthesized cross-domain image-label pairs. Formally, given a pair of non-mass and mass images and their corresponding malignancy labels, $(x_{nm}, y_{nm})$ and $(x_m, y_m)$, where $y \in \{0, 1\}$ represents benign and malignancy respectively, we generate a synthesized pair $(x_{nm}', y_{nm}')$ by:
\begin{equation}
    x_{nm}' = \lambda x_{nm} + (1 - \lambda)x_{m}
\end{equation}
\begin{equation}
    y_{nm}' = \lambda y_{nm} + (1 - \lambda)y_{m}
\end{equation}
where $\lambda \sim Beta(\alpha, \alpha)$. $\alpha$ is a hyperparameter (details in section~\ref{res}). Considering the training stability, we control the frequency of CrossMix with a hyperparameter $P=0.5$ with the insight of randomly maintaining original samples from different domains. Noted that our proposed CBN requires a domain label during training. For mixed image, $\delta(\lambda>0.5)$ is used as the domain label. 


Finally, with the above modules, the classification network $f_\theta(\cdot)$ is optimized by Cross Entropy(CE) loss:
\begin{equation}
    \theta ^{*} = \arg\min_{\theta} \mathcal{L}_{\text{CE}} (f_\theta(x_{nm}'), y_{nm}')
\end{equation}






\begin{table}[t]
\centering
\caption{Comparison of classification performance of non-mass data in \textbf{AUC} (mean $\pm$ std \%).}
\begin{tabular}{c|c|c|c}
\hline
Methods & \multicolumn{2}{c|}{Baseline} & \textbf{Ours}\\
\hline 
Datasets & Non-mass & Mixed & Mixed \\
\hline
Average & 68.97 $\pm$ 1.78 & 66.15 $\pm$ 3.48 & \textbf{76.47 $\pm$ 1.97} \\
\hline
\end{tabular}
\label{main_exp}
\end{table}

\begin{table}[tb]
\centering
\caption{Average classification performance in \textbf{AUC} (mean $\pm$ std \%) with different modules.}
\begin{tabular}{p{1.5cm} p{1.5cm} p{1.5cm} | c}
\hline
\multicolumn{3}{c|}{Modules} & Results\\
\hline
CBN & DA & CrossMix & Average\\
\hline
\checkmark & $\times$ & $\times$ & 72.64 $\pm$ 1.33\\
\checkmark & \checkmark & $\times$ & 74.16 $\pm$ 1.36\\
\checkmark & \checkmark & \checkmark & \textbf{76.47 $\pm$ 1.97}\\
\hline
\end{tabular}
\label{ab_1}
\end{table}

\begin{table*}[ht]
\centering
\caption{Average classification performance in \textbf{AUC} (mean $\pm$ std \%) with different hyperparameter $\alpha$ in CrossMix.}
\begin{tabular}{c|cccccc}
\hline
$\alpha$ & 0.1 & 0.2 & 0.3 & 0.4 & 0.5 & 0.6\\
\hline
AUC & 75.46 $\pm$ 1.47 & 75.50 $\pm$ 1.43 & 75.88 $\pm$ 1.60 & 76.07 $\pm$ 1.86 & \textbf{76.47 $\pm$ 1.97} & 75.93 $\pm$ 1.99\\
\hline
\end{tabular}
\label{ab_2}
\end{table*}

\begin{table*}[tb]
\centering
\caption{Average classification performance in \textbf{AUC} (mean $\pm$ std \%) with controlled training data ratio.}
\begin{tabular}{c|ccccc}
\hline
Ratio & 10\% & 30\% & 50\% & 70\% & 90\% \\
\hline
Baseline (Mixed) & 60.85 $\pm$ 2.99 & 61.74 $\pm$ 3.48 & 63.11 $\pm$ 3.16 & 64.37 $\pm$ 3.32 & 65.60 $\pm$ 3.72 \\
Ours & 63.36 $\pm$ 3.40 & 67.11 $\pm$ 2.66 & 69.18 $\pm$ 2.19 & 72.05 $\pm$ 1.92 & 74.89 $\pm$ 1.98 \\
\hline
\end{tabular}
\label{res:controlratio}
\end{table*}

\begin{table}[tb]
\centering
\caption{Average classification performance in \textbf{AUC} (mean $\pm$ std \%) with different CrossMix settings of the mixed images' malignancy.}
\begin{tabular}{c|c}
\hline
Setting & AUC \\
\hline
Across Malignancy & 74.64 $\pm$ 1.48 \\
Same Malignancy & \textbf{76.47 $\pm$ 1.97} \\
\hline
\end{tabular}
\label{res:crossmalignancy}
\end{table}

\section{Experiments and Results}
\subsection{Dataset and Implementation Details}
An in-house dataset with mass and non-mass gray-scale US images is used for model training and performance evaluation. This dataset consists of 3,679 mass images (1,487 benign and 2,192 malignant) and 503 non-mass images (303 benign and 200 malignant) with ground truth provided by the pathological result. For all experiments, we report the mean and variance of AUC performance for non-mass data on the Benign/Malignant binary classification task when training on 5 different folds of the dataset.


As for the DNN based classification model, we use a 10-layer ResNet~\cite{ResNet} as the backbone. The model is trained with AdamW~\cite{AdamW} optimizer with an initial learning rate of $1e-4$ for 150 epochs. The learning rate is decayed by a factor of $0.1$ after 100 epochs. For image augmentation, all images are randomly cropped and resized into $224 \times 224$ pixels, followed by random flipping and color jittering to prevent overfitting and enhance the diversity of training data. The training and testing pipelines are implemented with PyTorch~\cite{PyTorch} on a NVIDIA RTX 2080 Ti GPU.

\subsection{Classification Results}
\label{res}

The comparison of the Benign/Malignant classification performance in AUC is presented in Table~\ref{main_exp}. The result shows that directly training on mixed dataset leads to a 2.9\% drop in AUC compared to only training on the non-mass subset due to the input and output domain shift, while our transfer learning framework achieves an improvement of 7.5\% and 10.4\% in AUC compared to directly training on non-mass subset and mixed dataset, respectively. 


We present an ablation study to investigate the sufficiency and necessity of the proposed modules in our framework (i.e. CBN, DA, and CrossMix). As could be seen in Table~\ref{ab_1}, removing the CrossMix module degrades the AUC by 2.3\%, and further removing the DA module degrades the AUC by 1.6\%. This quantitative result implies the efficacy of every module in our framework. In addition, we investigate the sensitivity of the hyperparameter $\alpha$ in CrossMix, which controls the sharpness of the parameter $\lambda$. The experiment result is shown in Table~\ref{ab_2}. As could be seen, the best performance in AUC is achieved when $\alpha = 0.5$, which is the balance point between diversity and stability of CrossMix. 

\subsection{Exploratory Experiments}
To further explore the stability and generalizability of our proposed framework with even fewer training samples, we performed experiments with the controlled ratio of training data. As indicated in Table~\ref{res:controlratio}, our method achieves consistently superior performance compared to directly mixing non-mass and mass data.

We also investigate the effect of mixing images across malignancy (i.e., mixing the malignant image with the benign one). As could be seen in Table~\ref{res:crossmalignancy}, mixing images across malignancy degrades the performance by 1.8\%, which is reasonable considering the semantic ambiguity of mixing benign and malignant images.






\section{Conclusion}
In this paper, we propose a unified transfer learning framework to enhance the diagnosis performance of the DNN model on non-mass breast ultrasound images with the knowledge transferred from mass ones. Our work indicates the feasibility and importance of knowledge transfer in the diagnosis of non-mass breast lesions with ultrasound images. However, there are also limitations to this work. For example, we did not consider integrating the doppler ultrasound images, a traditional technique that is commonly equipped in US instruments for capturing blood supply signals of target lesions, with which the diagnosis performance is expected to be further enhanced~\cite{doppler1,doppler2}. In addition, the potential prognosis predictability of non-mass breast ultrasound images is unexplored. Nevertheless, we hope that our work could contribute to the future research of automatic pre-diagnosis and prognosis prediction on non-mass lesions with ultrasound images.
\\
\\
\textbf{Compliance with Ethical Standards:} All procedures performed in studies involving human participants were in accordance with the ethical standards of the institutional and/or national research committee and with the 1964 Helsinki declaration and its later amendments or comparable ethical standards.\\
\textbf{Acknowledgements:} This work was supported in part by National Natural Science Foundation of China 62171282, 111 project BP0719010, Shanghai Jiao Tong University Science and Technology Innovation Special Fund ZH2018ZDA17, and Shanghai Municipal Science and Technology Major Project (2021SHZDZX0102).




\bibliographystyle{IEEEbib}
\bibliography{strings,refs}

\end{document}